\documentstyle[12pt]{article}

\newcommand{\be}{\begin{equation}}
\newcommand{\ee}{\end{equation}}

\renewcommand{\ll}{{\cal L}}
\newcommand{\ppp}[1]{\frac{\partial \ll}{\partial #1}}
\newcommand{\ppx}[3]{\frac{\partial #1}{\partial X^{#2}_{#3}}}
\newcommand{\ppps}[1]{\frac{\partial \ll_{{\em string}}}{\partial #1}}
\newcommand{\pps}[1]{\frac{\partial \ll_{{\em string}}}{\partial
X^{\mu}_{#1}}}
\newcommand{\nnn}[2]{\nabla_{#1} \nabla_{#2} X^{\mu}}
\newcommand{\xx}[2]{X^{#1}_{#2}}

\begin{document}
\setlength{\unitlength}{0.25cm}
\thispagestyle{empty}

\hfill              \begin{tabular}{c} {\sf preprint} \\
                                        {\sf TPJU-28-93} \\
                                    \end{tabular}

\vspace{1.7cm}

\begin{center}
{\LARGE Classical Open String Models} \\
{\LARGE in 4-Dim Minkowski Spacetime}
\footnote{Supported in part by
the KBN under grant 2 P302 049 05}

\vspace{1.5cm}

{\large Pawe\l{} W\c{e}grzyn\footnote{
e-mail address: wegrzyn@ztc386a.if.uj.edu.pl} }
\vspace{0.5cm}

{ \normalsize \sl
Department of Field Theory \\
Institute of Physics, Jagellonian University \\
30-059 Cracow 16, Poland \\[0.1cm]}
\end{center}

\vfill
\begin{center}
{\sc Abstract}
\end{center}

\begin{quotation}

Classical bosonic open string models in four-dimensional Minkowski spacetime
are discussed. A special attention is paid to the choice of edge conditions,
which can follow consistently from the action principle. We consider
lagrangians that can depend on second order derivatives of worldsheet
coordinates. A revised interpretation of the variational problem for such
string theories is given. We derive a general form of a boundary term that can
be added to the open string action to control edge conditions and modify
conservation laws. An extended boundary problem for minimal surfaces is
examined. Following the treatment of this model in the geometric approach, we
obtain that classical open string states correspond to solutions of a
complex Liouville equation. In contrast to the Nambu-Goto case, the Liouville
potential is finite and constant at worldsheet boundaries. The phase part of
the potential defines topological sectors of solutions.
\end{quotation}
\vfill
{\sf December 1993} \hfill

\setcounter{page}{0}
\newpage
\setcounter{page}{1}
\vspace{0.8cm}
\textheight=18cm
\section{Introduction.}

There is a common conviction that in order to gain more insight into the
dynamical structure of QCD we need most likely to use some string
representation of this theory. It is suggested by topological nature of $1/N$
expansion \cite{tho}, area confinement law found in the strong coupling
lattice
expansion \cite{wil}, the success of dual models in description of Regge
phenomenology, the existence of flux-line solutions in confining gauge
theories
\cite{no,bbz}. More arguments are presented in recent reviews
\cite{gross,polch}.

In spite of numerous works, there is still a state of confusion about the
existence of an exact, or even approximate, stringy reformulation of 4-dim QCD
at all distance scales. Even at any specific scale, it is not evident what is
the adequate set of string variables and fields and how they correspond to QCD
gauge fields. Referring only to long distance scale, we usually adopt the
naive, but lucid, picture of flux tubes regime. A pair of quarks in the
confining phase is joined by a colour flux concentrated in a thin tube. If
these quarks are kept sufficiently far apart, the flux tube behaves like a
vibrating string. Using string variables as collective coordinates, one should
in principle find flux tube excitations by some quantization of the string
action. The question what kind of the string action should be employed to
represent the flux tube has yet to be answered. It is conceivable that to the
lowest order the action is just given by the Nambu-Goto action, which decribes
an infinitely thin relativistic bosonic string with constant energy per unit
length. As is well known, we cannot be satisfied with this first approximation
because of some unacceptable features of the quantized Nambu-Goto string.
Apart
from the problems with conformal anomaly outside the critical dimension or
tachyons and undesirable massless states in quantum spectra (which are
presumably less embarrassing at the long distance scale \cite{olesen}), all
standard quantizations give the incorrect number of degrees of freedom if we
confront it with QCD predictions \cite{polch2}.

Basically, there are two ways to modify the 4-dim Nambu-Goto action. In the
first approach, keeping the conformal invariance we can place additional
fields
on the worldsheet \cite{gliozzi,lewe}. The conformal anomaly can be saturated
due to the contribution of new conformal fields. Since we can hardly justify
the assumption that only massless degrees of freedom are important at the
hadronic scale, so the respecting of conformal symmetry is here rather a
compromise to make our theory mathematically tractable. The second kind of
modifications of the Nambu-Goto action, advocated in many papers (e.g.
\cite{poly}), is to introduce new action terms representing interactions
between transverse string modes. The fact that Regge trajectories derived
directly from fundamental quark models \cite{sim,mt,wkt} depart somewhat from
straight lines is a strong argument
that vibrating string modes cannot be considered as free. Next, some couplings
between these modes (short-distance interactions) would cause preferring
smooth
string worldsheets, leading to a well-defined quantum theory. Unfortunately,
all such string "self-interaction" terms involve higher order derivatives in
their lagrangians. Theories with higher order derivatives usually reveal
embarrassing pathological features, like lack of the energy bound, tachyons
already on the classical level, unitarity violation due to the presence of
negative norm states. Presumably, it means that one must regard any particular
effective theory of this type with a limited range of validity. From technical
point of view, such theories of strings are non-linear and cannot be
linearized
by a suit choice of gauge. Subsequently, a string cannot be described as an
infinite set of oscillators and there is no analogue of Virasoro algebra. The
conformal symmetry is usually spoiled. All that makes the evaluation of
physical observables technically difficult.

In this paper, we discuss possible modifications of the Nambu-Goto model (or
any other specific bosonic string model) by the change of boundary conditions
for open strings. This aspect is not well explored in literature, even though
the choice of boundary conditions can be crucial for defining relevant open
string models. Let us give some examples:

- Taking the usual hadronic string picture, we assume that quarks live only at
the opposite endpoints of the string and communicate through their couplings
to
the string between. Then, to some extent the choice of worldsheet boundary
conditions determines quarks trajectories. For instance, in the classical
Nambu-Goto model they are rather peculiar, being boosted periodic light-like
(null) curves. Undoubtedly, the unsolved problem how the quark masses and
quantum numbers (spin, color) couple to the string variables, partly lies in
the proper specification of string edge conditions.

- It is obvious that any internal symmetry of the worldsheet is necessarily
broken when worldsheet boundaries are included. Conformal transformations or
the full set of all reparametrizations are examples of that. Correspondingly,
conformal field theories defined on surfaces with boundaries are usually
endowed with only one copy of Virasoro algebra (instead of two, as for closed
surfaces). Recently \cite{lewe}, on the same basis the chiral symmetry
breaking
mechanism has been included in hadronic string models. This simple observation
that the existence of boundaries restricts the group of local worldsheet
symmetries, indicates that physical observables can essentially depend on
fields or currents evaluated on string boundaries.

- One of the straightforward calculations to test some open string model
against QCD expectations is to evaluate the static interquark potential.
Asympotically at the long distances scale, this potential is linear and its
slope can be related to the string tension. The first quantum corrections give
an universal Coulomb term \cite{luscher} (Casimir effect), being the function
of the number of worldsheet fields and of their boundary conditions. In an
approximation of flux-tube action by some conformal string theory, one can
represent boundary conditions by the set of relevant conformal operators
inserted at boundaries \cite{cardy}. The physical states are now constructed
with the help of both bulk and boundary operators. The Coulomb term depends on
the effective conformal anomaly \cite{isz}, being the the total conformal
anomaly diminished by the weight of the lowest state. This weight is sensitive
to the choice of boundary operators \cite{gliozzi}.

- The influence of worldsheet boundaries on critical string field theories has
been discussed in recent papers (see \cite{green,green2} and references
therein). In the framework of BRST formalism in the critical dimension, we can
consider either Neumann-type (e.g. standard Nambu-Goto edge equations) or
Dirichlet-type boundary conditions imposed on worldsheet coordinates. With
Dirichlet conditions, we have no physical open strings, but the closed-string
theory is radically modified, particularly the massless spectrum. Instead of
characteristic exponential fall-off of fixed-angle scattering amplitudes for
string models at high energies, we obtain for Dirichlet strings power-like
behaviour, like for parton models. We see here that the special type of
worldsheet boundaries, where these boundaries are mapped to single spacetime
points, implies that some point-like structure may appear at high energies.

In this work, we restrict ourselves to discuss open bosonic string models
defined by local lagrangians densities that can depend on second order
derivatives of worldsheet coordinates. In Section 2, we present general
formulas suitable to perform classical analysis of such string models. In
comparison with earlier works on this subject, a different interpretation of
the variational problem for string actions with second order derivatives is
given. Moreover, all derived classical formulas are explicitly covariant with
respect to reparametrization transformations. In Section 3, we derive a
general
form of a boundary term that can be added to the action, allowed by
requirements of Poincare and reparametrization invariances. Such a term can
modify edge conditions for open strings while bulk equations of motion are
preserved. Canonical conserved quantities are modified by some edge
contributions. Section 4 is devoted to the classical analysis of the string
model defined by the Nambu-Goto action with some new boundary terms added. It
is argued that such an open string model can be a suitable modification of the
Nambu-Goto model as far as  hadronic string interpretation is concerned. We
carry out the classical analysis using the geometric approach, which is
particularly convenient for our purposes. The classical open string
configurations that extremize the extended action correspond to solutions of a
complex Liouville equation. The relevant edge conditions for a Liouville field
are derived. The edge values are constant and finite there. Some preliminary
discussion about physical consequences is made. In Appendix, the notation used
throughout the paper is introduced and some basic mathematical definitions and
equations of surface theory are collected.

\section{String Lagrangians with Second Order Derivatives.}

In this section, we introduce some general formulas appertained to the
classical analysis of string models defined by lagrangians which depend on
second order derivatives of worldsheet radius vector. In comparison with
previous papers (e.g. \cite{nest,asw}), all formulas presented below are
explicitly
covariant with respect to the reparametrization, and we care especially with
the  correct derivation of edge conditions for open strings.

Let us consider the general form of the bosonic string action,
\be
S = \int_{\tau_{1}}^{\tau_{2}} d\tau \int_{0}^{\pi} d\sigma \ {\cal
L}_{string}  \ .
\label{s1}
\ee
It is convenient to represent the lagrangian density as
\be
\ll_{string} = \ll_{string}(X_{\mu,a};X_{\mu,ab}) = \sqrt{-g}
\ll(g^{ab};X_{\mu,a};\nabla_{a}\nabla_{b}X_{\mu}) \ ,
\ee
where $\ll$ is some scalar function made up of its specified arguments.
Having the string lagrangian with second order derivatives written down in the
above form, we can much easier perform mathematical calculations and keep the
explicit reparametrization invariance in all following steps.

To derive the classical equations of motion, we are to evaluate the
 variation of the string action under the infinitesimal change of the
worldsheet. Usually,  the following boundary conditions are assumed,
\be
\delta X_{\mu}(\tau_{i},\sigma) = \delta \dot{X}_{\mu}(\tau_{i},\sigma) = 0 \
,
\ \ i=1,2 \ .
\label{s3}
\ee
There is some subtle problem at this point. The above requirements suggest the
different interpretation of the variational problem in comparison with the
usual Nambu-Goto case. In (\ref{s3}), not only initial and final string
positions are fixed, but also the initial and final velocities of string
points. Therefore, if we consider some string at the time $\tau_{1}$, another
string at the time $\tau_{2}$ and some string trajectory being a solution of
Euler-Lagrange eqs. which interpolates between them, the solution does not
extremize the string action unless we restrict possible deviations of the
worldsheet to those that do not change its tangent vectors at the initial and
final positions. In other words, the string instant state is specified not
only
by its position, but also by its velocities. In fact, this modified
interpretation is not true as the boundary conditions (\ref{s3}) are not quite
proper for the string variational problem with second order derivatives. This
point will be clarified below.

The classical equations of motion following from (\ref{s1}) can be presented
in
the explicitly covariant form
\be
\sqrt{-g} \nabla_{a} \Pi^{a}_{\mu} = 0 \ ,
\label{s4}
\ee
where $\Pi^{a}_{\mu}$ is given by the following formula
\be
\Pi^{a}_{\mu} = - \ll \nabla^{a}X_{\mu} -  \ppp{X^{\mu}_{,a}} + 2 \ppp{g^{bc}}
g^{ab} \nabla^{c} X_{\mu} + \nabla_{b} \left[ \ppp{(\nnn{a}{b})} \right] \ .
\label{s5}
\ee
For open strings, the edge conditions at $\sigma=0,\pi$ must be satisfied,
\be
\sqrt{-g} \Pi^{1}_{\mu} + \partial_{0} \left[ \sqrt{-g} \ppp{(\nnn{0}{1})}
\right] = 0
\ ,
\label{s4a}
\ee

\be
\sqrt{-g}
\ppp{(\nnn{1}{1})} = 0 \ .
\label{s4b}
\ee
For the sake of more
convenient notation, here and throughout the paper we define and calculate the
variational derivatives of $\ll$ with the formal assumption that $g^{01}$ and
$g^{10}$, $\nnn{0}{1}$ and $\nnn{1}{0}$ are independent variables. Thus, all
variational derivatives on r.h.s. of (\ref{s5}) are tensor objects with
respect
to the reparametrization invariance. The covariance of edge conditions becomes
easy to check if we remind that in the presence of the worldsheet boundary any
reparametrization transformation $\sigma^{a} \rightarrow
\tilde{\sigma}^{a}(\tau,\sigma)$ must satisfy
\be
\tilde{\sigma}(\tau,0) = 0 \ \ , \ \
\tilde{\sigma}(\tau,\pi) = \pi \ .
\label{ss}
\ee
It is necessary in order to preserve the condition that the string parameter
$\sigma$ belongs to the interval $[0,\pi]$. In other case, performing the
variation of the string action we are forced to implement the variations due
to
the change of $\sigma$-interval, and the fact that the set of allowed
reparametrization transformations is restricted for open strings manifests in
additional Euler-Lagrange eqs.

The derivation of (\ref{s4}) from the standard Euler-Lagrange variational
equations is straightforward, so let us only cite the following identities
used
in this derivation
\be
\ppp{(\nnn{a}{b})} X^{\mu}_{,c} = 0 \ .
\label{s6}
\ee
To prove the above identities for lagrangians which include only scalar
constant parameters, it is enough to notice that the scalar
(with respect to both reparametrization and Poincare transformations) function
$\ll$ can be composed of the following "building blocks"

$$g^{ab} \ \ , \ \ \epsilon^{\mu \nu \rho \sigma}
(\nabla_{a}\nabla_{b}X_{\mu})
(\nabla_{c}\nabla_{d}X_{\nu})X_{\rho,e}X_{\sigma,f} \ \ , \ \ \nnn{a}{b}
\nabla_{c}\nabla_{d}X_{\mu} \ , $$

and refer to the trivial identities

$$ (\nnn{a}{b}) X_{\mu,c} = 0 \ . $$

In general, the origin of identities (\ref{s6}) lies in the reparametrization
invariance
of the string action (\ref{s1}). The full set of all Noether identities (see
(\ref{n1}-\ref{n2})) following from the reparametrization invariance of the
string action with second order derivatives has been derived in \cite{pw}.

Let us return to the problem of boundary conditions (\ref{s3}) imposed on the
variations of the worldsheet. If we assumed only that
\be
\delta X_{\mu}(\tau_{i},\sigma) = 0 \ , \ \ i=1,2 \ ,
\label{s7}
\ee
then using eqs. of motion (\ref{s4}) together with edge conditions
(\ref{s4a},\ref{s4b}) we would obtain the following result for the variation
of
the string action
\be
\delta S = \int_{0}^{\pi} d\sigma \ \sqrt{-g} \ppp{(\nnn{0}{0})} \delta
\dot{X}^{\mu} \Big|^{\tau=\tau_{2}}_{\tau=\tau_{1}} \ .
\label{s8}
\ee
If $g=0$ or the surface is locally flat then the following term vanishes, else
we can choose parametrization in such a way that the four vectors
($\dot{X}_{\mu},X'_{\mu},\ddot{X}_{\mu},\dot{X}'_{\mu}$) are linearly
independent at the point of the worldsheet with $\tau=\tau_{i}$. Then, we can
write down the general form of $\delta \dot{X}_{\mu}$ as the linear
combination
of these vectors,
\be
\delta\dot{X}_{\mu} = a_{1} \dot{X}_{\mu} + a_{2} X'_{\mu} + a_{3}
\ddot{X}_{\mu} + a_{4} \dot{X}'_{\mu} \ .
\label{s9}
\ee
On the other hand, the variation $\delta\dot{X}_{\mu}$ induced by the change
of
parametrization $\sigma^{a} \rightarrow \sigma^{a} + \delta\sigma^{a}$ is
given
by
\be
\delta\dot{X}_{\mu} = - \ddot{X}_{\mu} \delta\sigma^{0} - \dot{X}'_{\mu}
\delta\sigma^{1} \ .
\label{s10}
\ee
It means that the variations of $\dot{X}_{\mu}$ in the directions of
$\ddot{X}_{\mu}$ and $\dot{X}'_{\mu}$ are not important, because they can be
removed by the change of parametrization. In turn, if we restrict ourselves to
the
"physical" variations of the worldsheet, then with the help of identities
(\ref{s6}) we conclude that the term (\ref{s8}) vanishes.

Therefore, there are two ways to define properly the variational problem for
string action functionals which depend on second order derivatives. One way is
to assume boundary conditions (\ref{s7}) together with the additional
requirements that the variations $\delta\dot{X}_{\mu}$ in the directions of
$\ddot{X}_{\mu}$ and $\dot{X}'_{\mu}$ vanish, what in light of (\ref{s10})
means
that the choice of the parametrization of the worldsheet is locally fixed at
boundary points $\tau=\tau_{i}$. Other way is to take only the boundary
conditions (\ref{s7}), as in the Nambu-Goto case, and together with relevant
eqs. of motion and edge conditions we obtain additional equations
\be
\sqrt{-g} \ppp{(\nnn{0}{0})} = 0 \ \ \ {\rm  for} \ \ \tau=\tau_{1},\tau_{2} \
,
\ee
which have no dynamical content and impose only some boundary constraints on
the choice of worldsheet parametrization. Recapitulating, the interpretation
of
the variational problem for string actions with second order derivatives is
the
same as in the usual Nambu-Goto case. To derive the classical dynamics of
strings from the variational principle it is just enough to consider the
boundary conditions (\ref{s7}), i.e. to assume that the initial and final
string positions are fixed. The appearence of the term (\ref{s8}) in the
action
variation and resulted equations reflect only the fact that the
geometrical definitions
of the {\em initial} and {\em final} string positions are not invariant.

One more comment on the derivation of edge conditions should be made. They are
an integral part of equations of motion. They arise as in the variational
problem for open string worldsheets the whole boundary of the worldsheet is
not
fixed (like in an ordinary Plateau problem for two-dimensional surfaces), but
only a part of it composed of the initial and final string positions. The
other
part of the worldsheet boundary, defined by trajectories of string endpoints,
is not fixed (the ends of open strings are free). However, we can use another
equivalent method for the derivation of edge conditions. In the variational
problem we can dispense with considering the edge variations (assuming that
the
whole worldsheet boundary is fixed), and the edge conditions are produced when
we demand that there is no flow of the canonical Noether invariants through
the
string ends. In distinction with the Nambu-Goto case, for strings with second
order derivatives it is not enough to assure only that the canonical momentum
is conserved. We must check the same independently for the angular momentum,
because of its "spin part" induced by higher order derivatives. The comment on
the latter method of the edge conditions derivation is relevant to the recent
work of Boisseau and Letelier \cite{bl}. They make use of the internal
geometrical description of worldsheets to study models of strings with second
order derivatives. In this approach, they gain some new insight into the
content of dynamical equations. However, their formalism should be corrected
for open strings. The set of edge conditions derived from the conservation of
total energy-momentum should be supplemented by additional conditions
associated with the total angular momentum conservation. In particular, it
changes some results of the work \cite{bl}. For example, the prediction that
the endpoints of the Polyakov rigid string can travel with a speed less than
the velocity of light is not valid. Just taking into account the missing set
of
edge conditions, we check again that these velocities must be light-like, what
agrees with the independent proof of this fact given in \cite{pw}.

In the last part of this section, we write down formulas for Noether
invariants. The total momentum reads
\be
P_{\mu} = \int_{0}^{\pi} d\sigma \ p_{\mu} \ ,
\ee
where

$$p_{\mu} = - \ppps{X^{\mu}_{,0}} + \partial_{0} \left( \ppps{X^{\mu}_{,00}}
\right) = \sqrt{-g} \Pi^{0}_{\mu} - \partial_{1} \left[ \sqrt{-g}
\ppp{(\nnn{0}{1})} \right] \ . $$

The total angular momentum can be calculated from the following formula,
\be
M_{\mu\nu} = \int^{\pi}_{0} d\sigma \ m_{\mu\nu} \ ,
\ee
where

$$m_{\mu\nu} = x_{[\mu}p_{\nu]} + \ppps{X^{[\mu}_{,0a}} X_{\nu ],a} =$$

$$ = \sqrt{-g} X_{[\mu} \Pi^{0}_{\nu ]} - \sqrt{-g}
X_{[\mu,a} \ppp{(\nabla_{a} \nabla_{0} X^{\nu ]})} -
\partial_{1} \left[ \sqrt{-g} X_{[\mu} \ppp{(\nabla_{0} \nabla_{1} X^{\nu ]})}
\right] \ .
$$

\section{Boundary Terms for String Actions.}

We discuss the general string action functional with some boundary term added,
\be
S = \int d^{2}\sigma \ \ll^{{\rm bulk}}_{{\rm string}} - \int d^{2}\sigma \
\partial_{a} V^{a} \ .
\label{c1}
\ee
The stationarity of this action results in  some equations for the
interior of the string following from $\ll_{{\rm string}}^{{\rm bulk}}$ ,
and the role of the second action term is to ensure a
 more general set of edge conditions for an open string case.
Below, we will find the general form of this term allowed by requirements of
the locality, Poincare and reparametrization invariance. We restrict ourselves
to string lagrangians which depend on not higher than second order
derivatives,
what implies that
\be
\ppx{V^{a}}{\mu}{,bc} \xx{\mu}{,abc} = 0 \ .
\ee
The above identities give immediately the following equations
\be
\ppx{V^{0}}{\mu}{,00} = \ppx{V^{0}}{\mu}{,11} + 2 \ppx{V^{1}}{\mu}{,01} =
\ppx{V^{1}}{\mu}{11} = \ppx{V^{1}}{\mu}{,00} + 2 \ppx{V^{0}}{\mu}{,01} = 0 \ ,
\ee
and their general solution is of the form
\be
V^{a} = \epsilon^{ab} \tilde{A}^{c}_{\mu}\xx{\mu}{,bc}  + \tilde{B}^{a} \ ,
\label{c2}
\ee
where $\tilde{A}^{c}_{\mu}$ and   $ \tilde{B}^{a}$ are some arbitrary
functions
which depend on $X_{\mu}$ and their first derivatives. The translational
invariance of the action requires that
\be
0 = \ppx{(\partial_{a}V^{a})}{\mu}{} = \partial_{a} \left( \ppx{V^{a}}{\mu}{}
\right) \ ,
\ee
therefore there exists function $\Lambda_{\mu}(X_{\nu};X_{\nu,a})$ such that
\be
\ppx{V^{a}}{\mu}{} = \epsilon^{ab} \partial_{b} \Lambda_{\mu} =
\epsilon^{ab} \left( \ppx{\Lambda_{\mu}}{\nu}{} \xx{\nu}{,b}
+ \ppx{\Lambda_{\mu}}{\nu}{,c} \xx{\nu}{,bc} \right) \ .
\label{c3}
\ee
Comparing (\ref{c3}) with (\ref{c2}) we obtain
\be
\ppx{\Lambda_{\mu}}{\nu}{,a} = \ppx{\tilde{A}^{a}_{\nu}}{\mu}{} \ ,
\label{c4}
\ee
\be
\epsilon^{ab} \ppx{\Lambda_{\nu}}{\mu}{} \xx{\mu}{,b} =
\ppx{\tilde{B}^{a}}{\nu}{} \ .
\label{c5}
\ee
The above equations are consistent provided that
\be
\ppx{\Lambda_{\mu}}{\nu}{} - \ppx{\Lambda_{\nu}}{\mu}{} = F_{\mu\nu} \ ,
\ee
where $F_{\mu\nu}$ is some constant antisymmetric tensor. Consequently, there
exists a scalar function $\lambda(X_{\mu};X_{\mu,a})$ such that

$$\Lambda_{\mu} = \frac{1}{2} F_{\mu\nu} X^{\nu} + \ppx{\lambda}{\mu}{} \ . $$

Inserting this result in (\ref{c4},\ref{c5}), after some straightforward steps
we get the general form of $V^{a}$
\be
V^{a} = \epsilon^{ab} \partial_{a} \lambda  + \frac{1}{2} F_{\mu\nu} X^{\mu}
\xx{\nu}{,b}
+
\epsilon^{ab} A^{c}_{\mu} \xx{\mu}{,bc} + B^{a} \ ,
\label{c6}
\ee
where new arbitrary functions $A^{c}_{\mu}$ and $B^{a}$ depend now only on the
first derivatives of $X_{\mu}$. The first term on r.h.s. of (\ref{c6}) can be
omitted, as it does not contribute to $\partial_{a}V^{a}$.

The next step is to assure that the string action boundary term in (\ref{c1})
defined with the general functional $V^{a}$ of the form (\ref{c6}) is
reparametrization invariant. For this purpose, it is convenient to use the
Noether theorem for strings with second order derivatives, namely that the
string action functional is invariant under the reparametrization
transformations if and only if the lagrangian satisfies the following set of
identities \cite{pw}
\be
\pps{,ab} \xx{\mu}{,c} = 0 \ ,
\label{n1}
\ee
\be
\pps{,a} \xx{\mu}{,b} + \pps{,ad} \xx{\mu}{,bd} + \pps{,aa} \xx{\mu}{,ab} -
\ll_{string} \delta_{b}^{a} = 0 \ ,
\ee

$$\Big[ \pps{} - \partial_{d} \left( \pps{,d} \right) + \partial_{0}^{2}
\left(
\pps{,00} \right)$$
\be
 + \partial_{0} \partial_{1} \left( \pps{,01} \right) +
\partial_{1}^{2} \left( \pps{,11} \right) \Big] \xx{\mu}{,a} = 0 \ ,
\label{n2}
\ee
where fixed indices $a,b,c$ can take values 0 or 1 while the summation
over $d$ is assumed. Substituting (\ref{c6}) into the above equations, we end
up with some final general solution for $V^{a}$, which leads to the following
general form of the lagrangian density,
\be
\partial_{a} V^{a} = \frac{1}{2} \alpha \sqrt{-g} R + \beta \sqrt{-g} N +
\ll_{ext} \ ,
\label{c7}
\ee
where $\alpha$ and $\beta$ are some dimensionless constants, and $\ll_{ext}$
stands for boundary lagrangians which include Poincare vector or tensor
constants, i.e. describe some open systems with external fields. For such
lagrangians we have infinitely many possibilities, let us only give some
examples

$$\frac{1}{2} \epsilon^{ab} F_{\mu\nu} \xx{\mu}{,a} \xx{\nu}{,b} \ , $$

$$\sqrt{-g} A^{\mu} \Delta X_{\mu} \ ,$$

$$\sqrt{-g} \nabla_{a} \left( \frac{A^{\rho} \nabla^{a}
X_{\rho}}{\sqrt{1+ (T_{\mu\nu}
g^{bc} \xx{\mu}{,b} \xx{\nu}{,c})^{2}}} \right) \ , \ \ \ {\rm etc.}$$
The first term can be interpreted as the coupling of the charged string
endpoints with the external electromagnetic field \cite{bn}.

There are present only two string self-interaction terms  in
(\ref{c7}).
The relevant coefficients $B^{a}$ in (\ref{c6}) for these terms vanish, and
the
coefficients $A^{a}_{\mu}$ can be calculated from the equations
\be
\ppx{A^{a}_{\mu}}{\nu}{,b} - \ppx{A^{b}_{\nu}}{\mu}{,a} =
\frac{\alpha}{\sqrt{-g}}
\epsilon^{ab} G_{\mu\nu} + \beta g^{ab} \tilde{t}_{\mu\nu}  \ .
\ee
Note that the scalar density requirement on $\partial_{a} V^{a}$ does not
imply
that $V^{a}$ behaves  like a vector density under the reparametrization
transformations. The considered two self-interaction terms exemplify the case.

Let us summarize the results of this section. We proved that the generic local
term which can be added to any specific
string action to modify
edge conditions for open strings, provided that bulk equations of motion are
preserved, has the form  (\ref{c7}). We have obtained this conclusion
considering only Poincare and reparametrization invariance, and restricting
ourselves to local lagrangians with not higher than second derivatives. We did
not presume that this term should be polynomial or analytical in fields as
well
as no "power-counting" arguments for renormalizability of the quantized theory
were applied.  Thus, our result derived from a small set of very fundamental
assumptions has a general significance.

Remarkably, the only two self-interaction terms displayed on r.h.s. of
(\ref{c7}) are polynomial and well known in literature. In Euclidean
four-dimensional space, they are topological and related to Euler
characteristics and the numbers of self-intersections of two-dimensional
surfaces.

\section{Minimal Open String Models.}

In this section, we examine the string action functional for minimal
time-like surface models, defined by the following lagrangian
\be
\ll_{{\rm string}} = - \gamma \sqrt{-g} - \frac{1}{2} \alpha \sqrt{-g} R -
\beta
\sqrt{-g} N \ .
\label{f1}
\ee
The first term is the Nambu-Goto lagrangian, $\gamma$ stands for string
tension. The parameters $\alpha$ and $\beta$ are dimensionless. Let us also
introduce an        angle parameter $\theta \in [-\pi,\pi]$ defined as
\be
\tan{\frac{\theta}{2}} = \frac{\beta}{\alpha} \ .
\ee
According to the discussion in the previous section, the lagrangian (\ref{f1})
defines the most general model for free open strings, which worldsheets
represent minimal time-like surfaces of zero mean curvature.

Both new terms displayed on r.h.s. of (\ref{f1}) can be relevant for the
definition of the hadronic string action. The first boundary term is related
to
Euler characteristics in its Euclidean version. The genus factors that appear
in Polyakov quantum sum over surfaces \cite{poly2} can be interpreted as a
result of adding such a term to the string action. On the other hand, we will
show in this section that this selfinteraction term acts like "mass" term and
prevents string ends from propagating with light-like velocities. It may help
to couple consistently quark masses to hadronic strings. The relevance of the
second boundary term in (\ref{f1}) to QCD string has been also pointed out in
many papers. Polyakov \cite{poly} suggested that the inclusion of the term
that
weights worldsheets according to the number of selfintersections could assure
the existence of "smooth" phase of surfaces. In other works \cite{mn,re,jp},
this term has been used to reproduce an effect of QCD $\theta$-vacua in string
models. The exact correspondence between the moduli space of the maps
associated with a surface theory and the moduli space of the instanton sector
of QCD (or any other Yang-Mills theory) has been elaborated in \cite{vgjr}.
Exact instanton solutions in the string model with the selfintersections term
have been considered in \cite{gr}. Finally, in the paper \cite{jp} it is
argued
that this term is necessary for QCD string also in respect of having quark
spins included. Below, we will see on the classical level that the Minkowski
version of selfintersections term induces the topological sectors of
solutions,
what could correspond to the degenerated vacuum.

Equations of motion following from (\ref{f1}) are the same as in the usual
Nambu-Goto theory,
\be
\Delta X_{\mu} = 0 \ ,
\label{f2}
\ee
but supplementary edge equations for string endpoints $\sigma = 0,\pi$ are now
affected by additional terms, and have the following more general form
\be
\gamma \sqrt{-g} \nabla^{1} X_{\mu} - \alpha \partial_{0} \left(
\frac{1}{\sqrt{-g}} \nabla_{0} \nabla_{1} X_{\mu} \right) - \beta \partial_{0}
\left( \tilde{t}_{\mu\nu} \nabla^{0} \nabla_{0} X^{\nu} \right) = 0 \ ,
\label{f3}
\ee
\be
\frac{\alpha}{\sqrt{-g}} \nabla_{0} \nabla_{0} X_{\mu} - \beta
\tilde{t}_{\mu\nu} \nabla^{1} \nabla_{0} X^{\nu} = 0 \ .
\label{f4}
\ee
We will investigate the string dynamical problem given by the system of
equations (\ref{f2},\ref{f3},\ref{f4}). The best way is to use the geometrical
approach \cite{omn,bn}, i.e. to express the content of these equations in
terms
of
worldsheet curvature coefficients. Then, the differential equations transform
into algebraic ones. Eq.(\ref{f2}) says that the mean curvature is zero at any
point of the worldsheet, namely
\be
g^{ab} K^{i}_{ab} = 0 \ .
\label{f5}
\ee
Edge conditions (\ref{f3},\ref{f4}) can be integrated with respect to
worldsheet time $\tau$ and, after projections onto tangent and normal planes
respectively, they yield
\be
\frac{\alpha}{\sqrt{-g}} K^{i}_{00} + \beta \epsilon^{ij} K^{j1}_{0} = 0 \ ,
\label{f6}
\ee
\be
\frac{\alpha}{\sqrt{-g}} K^{i}_{01} - \beta \epsilon^{ij} K^{j0}_{0} = w^{i} \
,
\label{f7}
\ee
\be
\gamma \sqrt{-g} - w^{i} K^{i}_{01} = 0 \ ,
\label{f8}
\ee
where $w^{i}$ are arbitrary functions satisfying
\be
D_{0} w^{i} \equiv \partial_{0} w^{i} - \epsilon^{ij} \omega_{0} w^{j} = 0 \ .
\label{f9}
\ee
Let us choose one of the string endpoints, specified by $\sigma=0$ or
$\sigma=\pi$. We have here seven linear algebraic equations for local values
of
six curvature coefficients $K^{i}_{ab}$, so we can easy find that the solution
exists only if the following condition is satisfied
\be
\alpha w^{i} w^{i} = \gamma (\alpha^{2} + \beta^{2}) \ .
\label{f10}
\ee
{}From (\ref{f9}) follows that the expression $w^{i}w^{i}$ is time
independent,
what is compatible with the relation (\ref{f10}). Next, we see that the
classical solutions exist only for positive sign of $\alpha$,
\be
\alpha > 0 \ .
\ee
It is also interesting to note that the classical model defined by the
action composed only of the Nambu-Goto  and "self-interaction" terms
($\gamma,\beta
\neq 0; \alpha=0$) is inconsistent.

If the relation (\ref{f10}) is satisfied, then the edge values of the
curvature
cofficients are easy calculable from (\ref{f5}-\ref{f8}), namely
\be
K^{i}_{00} = \frac{\beta g g^{11} \epsilon^{ij} w^{j}}{\alpha^{2}+\beta^{2}} \
,
\label{f11}
\ee
\be
K^{i}_{01} = \frac{\sqrt{-g} (\alpha w^{i} + \beta \sqrt{-g} g^{01}
\epsilon^{ij} w^{j})}{\alpha^{2}+\beta^{2}} \ ,
\label{f12}
\ee
\be
K^{i}_{11} = - \frac{2\alpha \sqrt{-g} g^{01} w^{i} + \beta
[1+(\sqrt{-g}g^{01})^{2}] \epsilon^{ij} w^{j}}{(\alpha^{2}+
\beta^{2}) g^{11}} \ .
\label{f13}
\ee
One can verify that the formulas (\ref{f11}-\ref{f13}) are covariant with
respect to both the worldsheet reparametrization and local orthogonal rotation
transformations. The scalar functions $R$ and $N$ take the following constant
values at the boundary of the worldsheet,

$$\frac{R}{2} = \frac{\gamma}{\alpha}
\frac{\beta^{2}-\alpha^{2}}{\alpha^{2}+\beta^{2}}
 = -\frac{\gamma}{\alpha}
\cos{\theta}
\ ,$$

$$N = - \frac{2\beta\gamma}{\alpha^{2}+\beta^{2}}
 = - \frac{\gamma}{\alpha}
\sin{\theta}
\
.$$
Using the above results, one can check that the lagrangian density (\ref{f1})
vanishes at the string endpoints, what is a general feature of bosonic open
string models.

Now, let us turn into the investigation of  classical solutions, satisfying
equations of motion together with pertinent edge conditions. As usual, we
choose the conformal gauge,
\be
\dot{X}^{2} + X'^{2} = \dot{X} X' = 0 \ ,
\label{g1}
\ee
which makes equations (\ref{f2}) linear and the general solution reads
\be
X_{\mu}(\tau,\sigma) = X_{L\mu}(\tau+\sigma) + X_{R\mu}(\tau-\sigma)  \ .
\label{g2}
\ee
Let us denote,
\be
\ddot{X}^{2}_{L} = - \frac{1}{4} q_{+}^{2} \ \ , \ \
\ddot{X}^{2}_{R} = - \frac{1}{4} q_{-}^{2} \ \ , \ \
q_{\pm} = q_{\pm}(\tau \pm \sigma) \ \ , \ \ q_{\pm} \ge 0 \ .
\ee
Accordingly,
\be
(\ddot{X} \pm \dot{X}')^{2} = - (K^{i}_{00} \pm K^{i}_{01})
(K^{i}_{00} \pm K^{i}_{01}) = - q^{2}_{\pm} \ .
\ee
One can introduce new variables (we follow here \cite{bn}),

$$\sqrt{-g} = e^{-\phi} \ \ , \ \ \ \ \psi = \alpha_{+} - \alpha_{-} \ \ ,$$
\be
K^{1}_{00} \pm K^{1}_{01} = q_{\pm} \cos{\alpha_{\pm}} \ \ , \ \
K^{2}_{00} \pm K^{2}_{01} = q_{\pm} \sin{\alpha_{\pm}} \ .
\label{def}
\ee
In the geometrical approach, the role of dynamical equations play
Gauss-Peterson-Codazzi-Ricci equations (see Appendix), being the embedding
conditions for the worldsheet embedded in enveloping Minkowski spacetime.
Referring to (\ref{gauss},\ref{pc},\ref{ricci}), one can evaluate
\be
e^{\phi}(\ddot{\phi} - \phi'') = 2 e^{2\phi} q_{+} q_{-} \cos{\psi} \ ,
\label{g_}
\ee
\be
\dot{\alpha}_{\pm} + \omega_{0} = \pm (\alpha'_{\pm} + \omega_{1}) \ ,
\label{pc_}
\ee
\be
\omega'_{0} - \dot{\omega}_{1} = - e^{\phi} q_{+} q_{-} \sin{\psi} \ .
\label{r_}
\ee
Peterson-Codazzi eqs. (\ref{pc_}) allow us to eliminate torsion coefficients.
Two other equations have a nice geometrical interpretation. Gauss eq.
(\ref{g_}) relates the internal curvature scalar $R$ (l.h.s. of (\ref{g_})) to
the scalar build of the external curvature coefficients (r.h.s. of
(\ref{g_})).
The internal curvature scalar is build of the connections $\Gamma^{a}_{bc}$,
introduced for the tangent reper bundle with defined reparametrization
transformations. Thus, Gauss eq. (\ref{g_}) describes an immersion of the
tangent bundle. Similarly, l.h.s. of Ricci eq. (\ref{r_}) is a scalar
expression build of the connections $\epsilon^{ij} \omega_{a}$ defined on the
orthogonal reper bundle, endowed with local SO(2) transformations. Looking at
the r.h.s. of (\ref{r_}) (up to a constant it is equal to the scalar $N$), we
can interpret Ricci eq. as the immersion of
the orthogonal bundle. We see that Gauss and Ricci eqs. couple "internal" with
"external" geometry, describing immersions of tangent and orthogonal
two-dimensional reper bundles in 4-dim Minkowski space-time. Remarkably, both
scalars $R$ and $N$ constructed from disposable connections and displayed in
the immersion equations have been used in (\ref{f1}).

After eliminating the extrinsic torsion, Gauss and Ricci equations read
\be
\ddot{\phi} - \phi'' = 2 e^{\phi} q_{+} q_{-} \cos{\psi} \ ,
\ee
\be
\ddot{\psi} - \psi'' = 2 e^{\phi} q_{+} q_{-} \sin{\psi} \ .
\ee
The above equations can be written as one equation on a  complex function
$\Phi \equiv \phi + i \psi$ (see \cite{bn}),
\be
\ddot{\Phi} - \Phi'' = 2 q_{+} q_{-} e^{\Phi} \ .
\label{g3}
\ee
The gauge choice (\ref{g1}) leaves the residual symmetry,
\be
\tau \pm \sigma \rightarrow h_{\pm} (\tau \pm \sigma) \ ,
\label{g4}
\ee
where $h_{\pm}$ are arbitrary monotonic functions. Taking

$$h_{\pm} (\tau) = \int_{\tau_{0}}^{\tau} d\tau' \ q_{\pm}(\tau') \ ,$$
($h_{\pm}$ are monotonical due to $q_{\pm} \ge 0$), the equation (\ref{g3})
rewritten in the new variables (\ref{g4}) takes the standard form of Liouville
equation,
\be
\ddot{\Phi} - \Phi'' = 2 e^{\Phi}  \ .
\label{l}
\ee

As it has been proved, the classical Nambu-Goto dynamics (minimal surface
problem) reduces to a complex Liouville equation (\ref{l}). The functions
$q_{\pm}$ are arbitrary and their choice saturates the gauge freedom
associated
with the repa\-ra\-metri\-zation invariance. Unlike other gauge theories, in
the minimal string model the gauge can be completely fixed without breaking
the
Lorentz invariance. Obviously, the simplest gauge choice complementary to
(\ref{g1}) is
\be
q_{\pm} = 1 \ .
\label{g9}
\ee
Later, we will show that this gauge choice is also allowed when edge
conditions for worldsheets with boundaries are taken into account.  Here, let
us note that if we restrict ourselves
to reparametrization transformations which preserve worldsheet boundaries
(\ref{ss}), what means that

$$h_{+}(\tau) = h_{-}(\tau) = h_{+}(\tau-2\pi)+2\pi \ ,$$
then the gauge choice (\ref{g9}) is possible provided that
\be
q_{+}(\tau) = q_{-}(\tau) = q_{+}(\tau+2\pi) \ .
\label{g10}
\ee
Assuming (\ref{g9}), the general solution of (\ref{g3}) reads
\be
\Phi = \log{\left( \frac{-4 f'(\tau+\sigma)
g'(\tau-\sigma)}{\left[ f(\tau+\sigma) - g(\tau-\sigma)
\right]^{2}} \right) } \ ,
\label{g5}
\ee
where $f$ and $g$ are arbitrary complex functions (not necessary
single-valued,
only $\Phi$ should be single-valued). The function $\Phi$ is left invariant
when
$f$ and $g$ are changed by a modular transformation,
\be
f \rightarrow \frac{af+b}{cf+d} \ \ , \ \
g \rightarrow \frac{ag+b}{cg+d} \ \ , \ \ ad-bc=1 \ .
\label{g6}
\ee
It is helpful to know how to translate a given solution $\Phi$ of Liouville
eq.
 into the explicit radius-vector representation of the string worldsheet
$X_{\mu}(\tau,\sigma)$. In order to achieve it, we need to integrate
Gauss-Weingarten eqs. (\ref{gw}). For this purpose, it is convenient to
introduce the reference system composed of two real $k_{\mu}, l_{\mu}$ and one
complex $a_{\mu}$ null vectors,
\be
k^{2}=l^{2}=a^{2}=ka=la=0 \ , \ \ kl=-a\bar{a} = 2 \ .
\ee
As a result of the integration of Gauss-Weingarten eqs. we obtain (function
arguments are omitted)
\be
\dot{X}_{L\mu} = \frac{1}{4|f'|}(|f|^{2} k_{\mu} -
fa_{\mu}-\bar{f}\bar{a}_{\mu}
+ l_{\mu}) \ ,
\ee
\be
\dot{X}_{R\mu} = \frac{1}{4|g'|}(|g|^{2} k_{\mu} -
ga_{\mu}-\bar{g}\bar{a}_{\mu}
+ l_{\mu}) \ .
\ee
In particular, we can choose

$$k_{\mu}=(1,0,0,1) \ , \ l_{\mu}=(1,0,0,-1) \ , \ a_{\mu}=(0,1,i,0) \ .$$
Then,
\be
\dot{X}^{\mu}_{L} = \frac{1}{4|f'|}
(1+|f|^{2},f+\bar{f},i(f-\bar{f}),1-|f|^{2}) \ ,
\label{g12}
\ee
\be
\dot{X}^{\mu}_{R} = \frac{1}{4|g'|}
(1+|g|^{2},g+\bar{g},i(g-\bar{g}),1-|g|^{2}) \ .
\label{g13}
\ee
As it could be expected, the modular transformations (\ref{g6}) coincide
with Lorentz transformations of $X_{\mu}$. The integration of
Gauss-Weingarten eqs. gives also results for $n^{i}_{\mu}$ variables, namely
(here $\partial_{\pm} = \partial_{0} \pm \partial_{1}$)

$$ n^{1}_{\mu} + i n^{2}_{\mu} =$$

$$\frac{i}{e^{\phi}\sin{\theta}} \left[ \partial_{+} \left( e^{\phi}
\dot{X}_{L\mu} \right) e^{i\alpha_{-}} - \partial_{-} \left( e^{\phi}
\dot{X}_{R\mu} \right) e^{i\alpha_{+}} \right] \  .$$

Insofar, we have proved that Nambu-Goto eqs. together with complete
Poincare-invariant gauge-fixing conditions (\ref{g1}) and (\ref{g9}) are
equivalent to the problem defined by a complex Liouville eq. (without any
additional constraints).
To examine open strings case,  let us proceed with the derivation of boundary
conditions for Liouville complex
field $\Phi$ equivalent to edge conditions (\ref{f3},\ref{f4}) for
$X_{\mu}$ following from
the lagrangian (\ref{f1}). It is straightforward to convince ourselves that
the
edge conditions, see (\ref{f11}-\ref{f13}), are satisfied if and only if
\be
e^{-\phi} = \sqrt{\frac{\alpha}{\gamma}} q_{+} \ , {\rm for} \ \sigma=0,\pi \
,
\label{g7}
\ee
\be
\psi = \pi - \theta \ {\rm mod} \ 2\pi \ , {\rm for} \ \sigma=0,\pi \ ,
\ee
\be
\psi' = 0 \ , \ {\rm for} \ \sigma=0,\pi \ ,
\label{g8}
\ee
\be
q_{+}(\tau) =  q_{-}(\tau) \ , \ \  q_{+}(\tau+2\pi)= q_{+}(\tau) \ .
\label{gg9}
\ee
We see that the edge eqs. (\ref{gg9}) are exactly the same as the conditions
(\ref{g10}). It means that the gauge choice (\ref{g9}) is allowed for open
strings as well.

Summarizing, the classical open string equations following from the lagrangian
(\ref{f1}) are equivalent (in conformal gauge (\ref{g1}) supplemented by
complementary conditions (\ref{g9})) to complex Liouville equation
\be
\ddot{\Phi} - \Phi'' = 2 e^{\Phi} \ ,
\label{gg10}
\ee
with constant Dirichlet boundary conditions for the real part $\phi = {\rm Re}
\Phi$,
\be
e^{\phi} = \sqrt{\frac{\gamma}{\alpha}}
\ \ \ {\rm for} \
\sigma=0,\pi \ ,
\label{gg11}
\ee
and periodic boundary conditions for the imaginary part $\psi = {\rm Im}
\Phi$,
\be
\psi = \pi - \theta \ {\rm mod} \ 2\pi \ , \ \psi' = 0 \ \ \ {\rm for} \
\sigma=0,\pi
\ .
\label{gg12}
\ee

We have evaluated the general form of boundary conditions which can follow
consistently from the string action for isolated open strings with not higher
than second derivatives. In this paper, we do not develope the thoroughgoing
analysis of classical string states. We restrict ourselves to indicate that
some essential differences appear while we are comparing the above defined
 extended boundary problem for minimal worldsheets with the ordinary
Nambu-Goto case.

First, let us consider the case $\beta = 0$ ($\theta = 0$). The lowest state
solution of Nambu-Goto model, that corresponds to stationary (soliton)
solution
for Liouville field, represents the rotating rigid rod,
\be
X_{\mu} = \frac{1}{\lambda^{2}}(\lambda \tau, \cos{(\lambda \tau)}
\sin{[\lambda (\sigma - \frac{\pi}{2})]}, \sin{(\lambda \tau)}
\sin{[\lambda (\sigma - \frac{\pi}{2})]},0) \ .
\ee
It is also a solution of our extended boundary problem
(\ref{gg10}-\ref{gg12}),
but now the string end points are no longer forced to travel with light-like
velocities. The parameter
$\lambda$ is subject to the following equation (for Nambu-Goto configurations
$\lambda=1$),
\be
\left[ \frac{\lambda}{ \cos{(\lambda \pi /2)}} \right]^{2} =
\sqrt{ \frac{\gamma}{\alpha} } \ .
\ee
For $\beta \neq 0$ ($\theta \neq 0$), there are no solitonic solutions of the
Liouville equation. The imaginary part of Liouville field $\psi$, that is an
angle variable (see definition (\ref{def})),
cannot be trivial (i.e. to be constant everywhere on the string). The mapping
$e^{i\psi}:[0,\pi]\rightarrow S^{1}$ provide us with some topological winding
number, classifying possible solutions.

In the end of this section, let us make a comment on the possible extension of
string action (\ref{f1}) by adding the rigidity term \cite{poly},
\be
{\cal L}_{\em rig} = \kappa \sqrt{-g} (\Delta X_{\mu})^{2} \ .
\ee
Obviously, the extended boundary problem for rigid strings is much more
complicated. However, it is nice to note that all classical open string
solutions defined by the system (\ref{gg10}-\ref{gg12}) are still exact
solutions when the extended boundary problem is formulated with the rigidity
action term taken into account.
Moreover, all these solutions carry the same energy and angular momentum in
both models (the rigidity term does not influence conservation laws for this
class of solutions). Presumably, these are the only open rigid string
solutions
around which a sensible semiclassical quantization can be performed.

\vspace{0.8cm}
\noindent {\bf Appendix}
\vspace{0.2cm}

In the appendix, we introduce notation and gather mathematical
equations of surfaces theory used throughout this paper. The string worldsheet
is denoted by $X_{\mu}(\sigma^{a})=X_{\mu}(\tau,\sigma)$ ($\sigma \in
[0,\pi]$), its derivatives either by $X_{\mu,a}$ ($a=0,1$) or by the dot and
the prime for the derivatives over $\tau$ and $\sigma$ parameters
respectively.
The following causality conditions are imposed on the worldsheet,

$$\dot{X}^{2} \ge 0 \ \ , \ \ \dot{X}_{0} > 0 \ \ , \ \ X'^{2} \le 0 \ .$$

The induced metric is $g_{ab}=X^{\mu}_{,a}X_{\mu,b}$, its determinant $g$
($g\le 0$). Christoffel coefficients $\Gamma^{a}_{bc}$, covariant
derivative $\nabla_{a}$ and raising and lowering indices (denoted by small
first roman letters ($a,b,c,...$)) are defined with respect
to the induced metric. The  Riemann-Christoffel tensor
$R_{abcd}$ is also defined as usual,

$$R_{abcd} = \partial_{c} \Gamma_{bda} - \partial_{d} \Gamma_{bca} +
\Gamma^{e}_{bc} \Gamma_{ade} - \Gamma^{e}_{bd} \Gamma_{ace} \ ,$$
and the internal curvature scalar $R$ is introduced together with the
following relation,

$$R^{c}_{\ acb} = \frac{1}{2}  g_{ab} R \ .$$

 At any point of the worldsheet two
orthonormal vectors
$n^{i}_{\mu}$
($i=1,2$) can be introduced,

$$n^{i}_{\mu} X^{\mu}_{,a} = 0 \ \ , \ \ \ n^{i}_{\mu} n^{j\mu} = -
\delta^{ij}
\ .$$

$$\frac{1}{\sqrt{-g}} \in_{\mu \nu \rho \sigma} \dot{X}^{\mu} X'^{\nu}
n^{1\rho} n^{2\sigma} = + 1 \ . $$

The last condition fixes the orientation of the local frame. There is still
some arbitrariness in a choice of orthonormal vectors $n^{i}_{\mu}$, namely
one
can perform a local SO(2)-rotation in a normal plane ($M$ stands for the
rotation matrix about the angle $\phi$),

$$n^{i}_{\mu} \rightarrow M^{ij}(\phi) n^{j}_{\mu} \ \ , \ \ \
\phi=\phi(\tau,\sigma) \ ,$$

$$\omega_{a} \rightarrow \omega_{a} + \partial_{a} \phi \ .$$

This freedom can be considered as a local symmetry of the system desribed in
the geometric approach. Therefore, for practical purposes it is convenient to
use "double-covariant" derivative $D_{a}$, i.e. the derivative covariant with
respect to both reparametrization change and local orthogonal rotation. This
derivative is defined with the help of respective connections
$\Gamma^{c}_{ab}$
and $\in^{ij} \omega_{a}$.

The projection operator onto the normal plane is denoted by $G_{\mu\nu}$,

$$G_{\mu\nu} = \eta_{\mu\nu} - g^{ab} X_{\mu,a} X_{\nu,b} \ ,$$

and antisymmetric tensor $t_{\mu\nu}$ is introduced as usual,

$$t_{\mu\nu} = \frac{1}{\sqrt{-g}} \in^{ab} X_{\mu,a} X_{\nu,b} \ \ \ , \ \
\tilde{t}^{\mu\nu} = \frac{1}{2} \in^{\mu\nu\rho\sigma} t_{\rho\sigma} \ .$$

Let us also define covariant tensor $N_{abcd}$ as

$$N_{abcd} = \tilde{t}^{\mu\nu} (\nabla_{a}\nabla_{b} X_{\mu})
\nabla_{c}\nabla_{d} X_{\nu} \ ,$$

and the scalar function $N$ together with the following relation,

$$N^{c}_{\ acb} = \sqrt{-g} \in_{ab} N \ .$$

The external curvature $K^{i}_{ab}$ and torsion $\omega_{a}$ coefficients are
defined with Gauss-Weingarten equations (in parantheses we give their form in
"double-covariant" notation),

$$X_{\mu,ab} = \Gamma^{c}_{ab} X_{\mu,c} + K^{i}_{ab} n^{i}_{\mu} \
, \ \ (D_{a}D^{b}X_{\mu}=K^{ib}_{a}n^{i}_{\mu} \ ,)$$

\be
\partial_{a} n^{i}_{\mu} = K^{i b}_{a} X_{\mu,b} + \epsilon^{ij} \omega_{a}
n^{j}_{\mu} \ , \ \ (D_{a}D^{b}n^{i}_{\mu}=K^{ib}_{a}X_{\mu,b} \ .)
\label{gw}
\ee
Instead of using radius vector coordinates, we can represent the surface (up
to
Poincare transformations) by induced metric and external curvature and torsion
coefficients, which satisfy the following identities (being the compatibility
conditions for Gauss-Weingarten eqs.)

\be
R_{abcd}=K^{i}_{ad}K^{i}_{bc}-K^{i}_{ac}K^{i}_{bd} \ ,
\ \ \ {\it (Gauss \ eqs.)}
\label{gauss}
\ee
\be
\nabla_{a}K^{i}_{bc}-\nabla_{b}K^{i}_{ac}=\epsilon^{ij}(\omega_{a}K^{j}_{bc}-
\omega_{b}K^{j}_{ac}) \ ,
\ \ {\it (Peterson-Codazzi \ eqs.)}
\label{pc}
\ee
\be
\partial_{a} \omega_{b} - \partial_{b} \omega_{a} = \epsilon^{ij} g^{cd}
K^{i}_{ac} K^{j}_{bd} \ .
\ \ {\it (Ricci \ eqs.)}
\label{ricci}
\ee
All above equations are covariant with respect to both reparametrization
change
and local orthonormal rotation. The Peterson-Codazzi eqs. and Ricci eqs. in
"double-covariant" notation have the following form,

$$ D_{a}K^{i}_{bc}=D_{b}K^{i}_{ac} \ ,$$

$$[D_{a},D_{b}] K^{i}_{cd} = - \sqrt{-g} \in_{ab} \in^{ij} N K^{j}_{cd} \ .$$

\end{document}